# The receptron is a nonlinear threshold logic gate with intrinsic multi-dimensional selective capabilities for analog inputs

B. Paroli[1,2*], F. Borghi[1,2], M.A.C. Potenza[1,2], P. Milani[1,2*]

[1]Dipartimento di Fisica "Aldo Pontremoli", Università degli Studi di Milano, Via Celoria 16, 20133 Milano, Italy

[2]Gruppo Cibernetici Operativi (GRUCIO), Via Celoria 16, 20133 Milano, Italy

*corresponding authors: bruno.paroli@unimi.it; paolo.milani@mi.infn.it

## Abstract

Threshold logic gates (TLGs) have been proposed as artificial counterparts of biological neurons with classification capabilities based on a linear predictor function combining a set of weights with the feature vector. The linearity of TLGs limits their classification capabilities requiring the use of networks for the accomplishment of complex tasks. A generalization of the TLG model called "receptron", characterized by input-dependent weight functions allows for a significant enhancement of classification performances even with the use of a single unit. Here we formally demonstrate that a receptron, characterized by nonlinear input-dependent weight functions, exhibit intrinsic selective activation properties for analog inputs, when the input vector is within cubic domains in a 3D space. The proposed model can be extended to the n-dimensional case for multidimensional applications. Our results suggest that receptron-based networks can represent a new class of devices capable to manage a large number of analog inputs, for edge applications requiring high selectivity and classification capabilities without the burden of complex training.

## 1. Introduction

A Threshold Logic Gate (TLG) is a mathematical model of the McCulloch and Pitts neuron consisting of an algorithm for supervised learning of binary classifiers [1]. The classification is based on a linear predictor function combining a set of weights with the feature vector. A practical software implementation of the TLG is the Perceptron: the elemental building block of artificial neural networks (ANNs) [2]. A single-layer perceptron can classify only linear separable boolean functions, multiple-layer perceptrons are required to perform complex tasks such as pattern recognition [3], image classification [4,5], speech and text recognition [6,7], robotic control [8,9], and they are the backbone of current artificial intelligence technologies. The functional non-completeness of a single-layer perceptron limits its classification and data processing capabilities, thus larger networks are necessary to address demanding computing performances, making the network extremely complex with huge training times and energy consumption [10].

Recently we proposed a generalization of the perceptron model called "receptron", characterized by input-dependent weight functions [11-15]. This provides a significant enhancement of classification capabilities even within single layer schemes; the hardware implementation of the receptron scheme has been demonstrated both with electronic [15] and optical devices [13].

Here we formally demonstrate how a receptron, characterized by nonlinear input-dependent weight functions, exhibit intrinsic selective activation properties when the input vector is within cubic domains in a 3D space (for the sake of exemplification we consider a 3-input receptron). The proposed model can be extended to the n-dimensional case for multidimensional control applications. Our model can be implemented by two different and equivalent hardware schemes.

## 2. The receptron model

The receptron is a generalization of the McCulloch - Pitts [1] and Rosenblatt perceptrons [16] where the weights $w(\vec{x})$ are, more in general, functions of the inputs vector $\vec{x} = (x_1, \ldots, x_n)$ [13]. The weight functions and inputs are combined as

$$S(\vec{x}) = \sum_{j=1}^{n} x_j w_j(\vec{x}). \quad (1)$$

The receptron is activated, through the thresholding process

$$Y(\vec{x}) = \begin{cases} 1 & S > t_s \\ 0 & S \leq t_s \end{cases}, \quad (2)$$

where $t_s$ is a threshold parameter and $Y$ is the receptron output.

The thresholding process can be extended as in [13] by considering two threshold parameters $t_h$ and $t_l$ as

$$Y(\vec{x}) = \begin{cases} 1 & t_l < S \leq t_h \\ 0 & otherwise \end{cases}. \quad (3)$$

For a finite value of $S$, Eq. (3) is a more general case with respect to Eq. (2) since the former activation can be formally obtained from the latter by posing $t_h = +\infty$ and $t_s = t_l$.

The input-dependent weights enable a significantly higher computational capacity per individual device compared to fixed weights. In the case of receptrons with 4 digital inputs, functional completeness has been experimentally demonstrated [13], meaning that all 65,536 possible functions can be successfully computed using a single receptron. In contrast, a single perceptron would be limited to solving only 1,882 Boolean functions, achieving an efficiency of less than 3%.

For analog inputs the input-dependent weights provide selective properties at the single device level as proven in the next section.

## 3. Receptron selectivity in a real 3D space

To make a device output active when three inputs $x, y, z$ have values within a cubic domain (in a 3D space) centered in $(x_0, y_0, z_0)$ with volume $\Delta^3$ as sketched in Fig. 1, we introduce the overall activation function

$$Y = rect(x - x_0 / \Delta) \cdot rect(y - y_0 / \Delta) \cdot rect(z - z_0 / \Delta) \quad . \quad (4)$$

We find the weight functions of a receptron, if any, for the output activation $Y$. For simplicity and without loss of generality, we exclude the boundaries $\frac{x-x_0}{\Delta} \neq \frac{1}{2}$, $\frac{y-y_0}{\Delta} \neq \frac{1}{2}$ and $\frac{z-z_0}{\Delta} \neq \frac{1}{2}$.

By introducing $f_x = rect(x - x_0/\Delta)$, $f_y = rect(y - y_0/\Delta)$, $f_z = rect(z - z_0/\Delta)$ and exploiting the De Morgan theorem [17] we write

$$Y = f_x \cdot f_y \cdot f_z = \overline{\overline{f_x \cdot f_y \cdot f_z}} = \overline{\overline{f_x} \vee \overline{f_y} \vee \overline{f_z}} \quad . \quad (5)$$

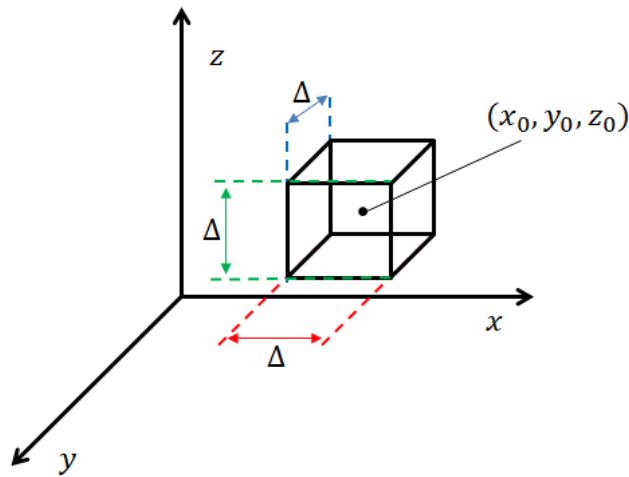

**Fig. 1** Sketch of the cubic domain due to the product of rectangular functions.

The logic operator in Eq. (5)

$$f_i \vee f_j \quad (6)$$

can be formally replaced by the algebraic sum $f_i + f_j$ properly normalized

$$f_i \vee f_j = \frac{f_i + f_j}{f_i + f_j + \overline{f_i} \cdot \overline{f_j}} \quad . \quad (7)$$

In table 1 we prove the equivalence given in Eq. (7) through the truth tables of the partial operation in Eq. (7).

| $f_i$ | $f_j$ | $f_i \vee f_j$ | $f_i + f_j$ | $f_i + f_j + \overline{f_i} \cdot \overline{f_j}$ | $\dfrac{f_i + f_j}{f_i + f_j + \overline{f_i} \cdot \overline{f_j}}$ |
|---|---|---|---|---|---|
| 0 | 0 | **0** | 0 | 1 | **0** |
| 0 | 1 | **1** | 1 | 1 | **1** |
| 1 | 0 | **1** | 1 | 1 | **1** |
| 1 | 1 | **1** | 2 | 2 | **1** |

**Table 1.** Truth table of the quantities defined in Eq. (7).

By substituting Eq. (7) into Eq. (5) we find

$$Y = \overline{\left(\frac{\overline{f_x} + \overline{f_y} + \overline{f_z}}{\overline{f_x} + \overline{f_y} + \overline{f_z} + f_x \cdot f_y \cdot f_z}\right)}, \quad (8)$$

that is written as

$$\overline{Y} = \frac{1}{D(\vec{x})} \left( \overline{f_x}(x) + \overline{f_y}(y) + \overline{f_z}(z) \right), \quad (9)$$

where $D(\vec{x}) = \overline{f_x} + \overline{f_y} + \overline{f_z} + f_x \cdot f_y \cdot f_z$ or equivalently to

$$\overline{Y} = \frac{1}{D(\vec{x})} \left( 1 - f_x(x) + 1 - f_y(y) + 1 - f_z(z) \right), \quad (10)$$

where the formula $\overline{f_x}(x) = 1 - f_x(x)$ has been used. Finally, we write Eq. (10) in compact form as

$$\overline{Y} = \frac{N}{D(\vec{x})}, \quad (11)$$

where $N = 1 - f_x(x) + 1 - f_y(y) + 1 - f_z(z)$.

Figure 2 shows the relationship between $\bar{Y}$ and $N$ (black squares). We observe that $N \geq 1$ when $\bar{Y} = 1$, while $\bar{Y} = 0$ implies $N = 0$, $D(\vec{x})$ acts as a normalization factor. Notice that the black squares are well described by a step function with threshold $t$, where $0 < t \leq 1$, therefore Eq. (11) is equivalent to

$$\bar{Y} = h(N - t), \quad (12)$$

where $h(.)$ is the Heaviside function. Making $N$ explicit, we write

$$\bar{Y} = h\big(1 - f_x(x) + 1 - f_y(y) + 1 - f_z(z) - t\big). \quad (13)$$

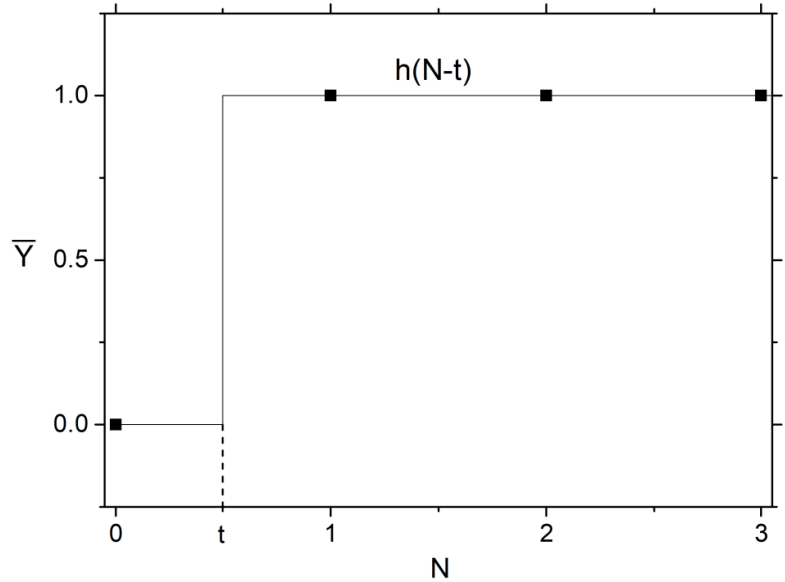

**Fig. 2** Graphic representation of the equivalence between Eq. (11) and the Heaviside function $h(N - t)$.

By negating Eq. (13) we obtain the result for the receptron representation

$$Y = \bar{h}\big(1 - f_x(x) + 1 - f_y(y) + 1 - f_z(z) - t\big), \quad (14)$$

which is formally the double threshold receptron shown in Eq.s (1-3) with weight functions

$$w_x(x) = \frac{1 - f_x(x)}{x}, \quad w_y(y) = \frac{1 - f_y(y)}{y}, \quad w_z(z) = \frac{1 - f_z(z)}{z}, \quad (15)$$

and thresholds $t_L < 0$, $t_h = t$.

These results show how the receptron model exhibits selective properties when the weights are nonlinear input-dependent functions as in Eq. (15). More precisely, by substituting Eq. (14) into Eq. (4), and considering the result of Eq. (15) we obtain

$$rect\left(\frac{x-x_0}{\Delta}\right) \cdot rect\left(\frac{y-y_0}{\Delta}\right) \cdot rect\left(\frac{z-z_0}{\Delta}\right) = \bar{h}(w_x(x)\,x + w_y(y)\,y + w_z(z)\,z - t) \quad (16)$$

The left part of Eq. (16) is formally a selective activation function in a cubic domain, which is equivalent to the right part of Eq. (16), i.e. a receptron output with nonlinear weights (see Eq. (15)). Consequently, a single-layer network of receptrons can exhibit selectivity. In contrast, a single-layer neural network composed of perceptrons lacks this selective capability.

## 4. Generalization to the n-dimensional space

For the n-dimensional space the activation domain is the hypercube

$$Y = f_1 \cdot f_2 \cdot \ldots \cdot f_n = \overline{\overline{f_1} \cdot \overline{f_2} \cdot \ldots \cdot \overline{f_n}} = \overline{\overline{f_1} \vee \overline{f_2} \vee \ldots \vee \overline{f_n}}. \quad (17)$$

Equation (7) can be generalized as

$$f_1 \vee f_2 \ldots \vee f_n = \frac{\sum_{i=1}^{n} f_i}{\sum_{i=1}^{n} f_i + \prod_{i=1}^{n} \overline{f_i}}. \quad (18)$$

Substituting Eq. (18) into Eq. (17) we obtain

$$Y = \overline{\frac{\sum_{i=1}^{n} \overline{f_i}}{\sum_{i=1}^{n} \overline{f_i} + \prod_{i=1}^{n} f_i}} \quad (19)$$

and we find again the ratio as in Eq. (11)

$$\bar{Y} = \frac{N}{D(\vec{x})}, \quad (20)$$

where $N = \sum_{i=1}^{n} 1 - f_i(x_i)$ and $D = \sum_{i=1}^{n} \overline{f_i} + \prod_{i=1}^{n} f_i$.

Using the same considerations as in the previous section we obtain

$$Y = \bar{h}(N - t), \quad (21)$$

or

$$Y = \bar{h}\left[\left(\sum_{i=1}^{n} 1 - f_i(x_i)\right) - t\right], \quad (22)$$

i.e. the receptron shown in Eq.s (1-3) with weight functions

$$w_i(x_i) = \frac{1 - f_i(x_i)}{x_i}, \quad (23)$$

and thresholds $t_L < 0$, $t_h = t$.

## 5. Discussion

In the previous sections we proved that the selective activation of an output $Y$ in a n-dimensional space, defined through the product of rectangular functions, is equivalent to the activation of a n-input receptron with weights $w_i(x_i) = \frac{1-f_i(x_i)}{x_i}$. In terms of hardware, this equivalence can be schematically represented (for $n = 3$) as in Fig. 3.

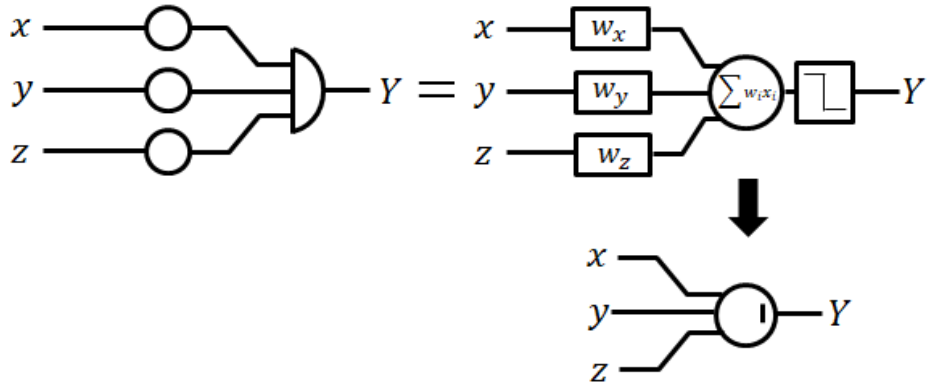

**Fig. 3** Equivalent hardware schemes for a selective device in a 3D space. Left: scheme (explicit) that exploits three individual devices (circles) implementing rectangular functions and a AND logic gate to combine the outputs. Right: scheme (implicit) realized with a single receptron with weight functions as in Eq. (15). Bottom: symbol proposed for a selective receptron.

While the first scheme exploits three single-input devices and a AND logic gate i.e. the implementation of the left part of Eq. (16), the second scheme exploits a single receptron with specific weight functions i.e. the implementation of the right part of Eq. (16). Although the two schemes are equivalent, they differ fundamentally in concept. The first scheme explicitly implements a selective device using the AND operator, this solution differs from the summation and the synaptic integration, mechanisms characteristic of neural biological systems [18-22]. On the other hand, the implicit scheme describes a selective device relying on the summation of stimuli followed by an activation function, provided the stimuli are properly weighted by input-dependent weight functions. In Fig. 4 we plot an example of the weight functions for an activation cubic domain centered in $(x_0, y_0, z_0) = (5,3,10)$. By changing the weight functions, we change the domain size $\Delta$ and shift the domain in the whole 3D space. Notice that the proof in the previous section does not loose validity when considering the more general case

$$Y = rect\left(x - \frac{x_0}{\Delta_x}\right) \cdot rect\left(y - \frac{y_0}{\Delta_y}\right) \cdot rect\left(z - \frac{z_0}{\Delta_z}\right), \qquad (24)$$

with $f_x = rect(x - x_0/\Delta_x)$, $f_y = rect(y - y_0/\Delta_y)$, $f_z = rect(z - z_0/\Delta_z)$, that describe a more general parallelepiped domain in a 3D space.

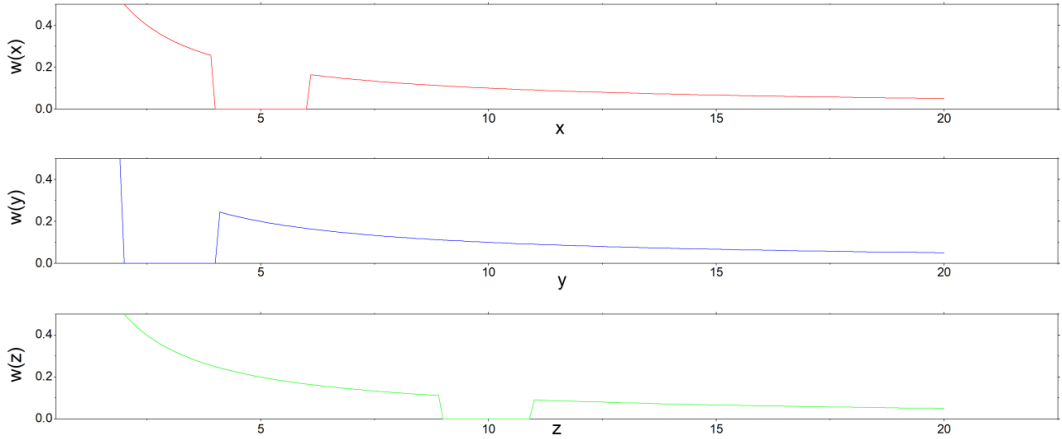

**Fig. 4** Example of weight functions for a selective receptron with activation conditions in the cubic domain centered in $(x_0, y_0, z_0) = (5,3,10)$ and with side $\Delta= 2$.

Moreover, the explicit representation in Eq. (4) can be generalized to any arbitrary logic function $F(f_x, f_y, f_z)$ (not necessarily AND operators), where F is one of the $2^n$ possible Boolean functions. This increases the logic relationships between the sub-spaces of the domains defined by $f_x, f_y, f_z$. An example for $(f_x \vee f_y) \wedge f_z$ is shown in Fig. 5.

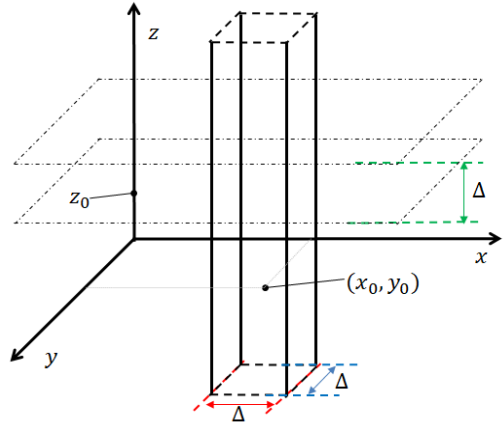

**Fig. 5** Selective properties of a receptron in open domains for $F = (f_x \vee f_y) \wedge f_z$.

Activation occurs when $x, y$ are within the vertical column with section $\Delta^2$ centered in $(x_0, y_0)$, or when $z$ is between the planes $z = z_0 + \Delta/2$ and $z = z_0 - \Delta/2$. This corresponds to the logic disjunction operation for activation conditions in open domains.

The logic disjunction operation for activation conditions in closed domains (e.g. cubic domains), can be realized by combining two or more selective receptrons in a two-layer network exploiting the OR operation as shown in Fig. 6. Here the network is composed of two selective receptrons (blue receptrons in Fig. 6) and a digital input receptron implementing the OR logic connective (red receptron), or by using a single multidimensional (6-inputs) receptron in a single layer, which is capable of solving the function $F = (f_1 \cdot f_2 \cdot f_3) \vee (f_4 \cdot f_5 \cdot f_6)$ and connected as shown in Fig. 7 (center).

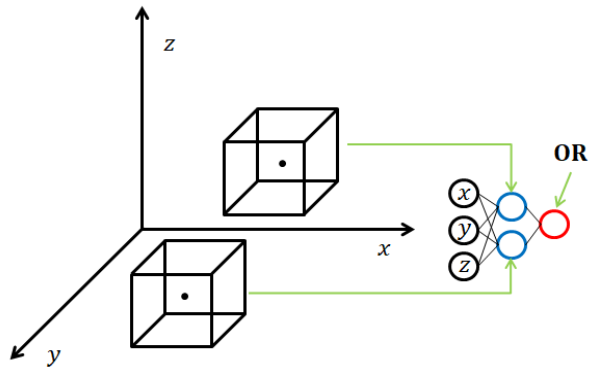

**Fig. 6.** Logic disjunction operation for the activation conditions in closed domains (two cubic domains) realized with a network of selective receptrons as in Fig. 3 (right).

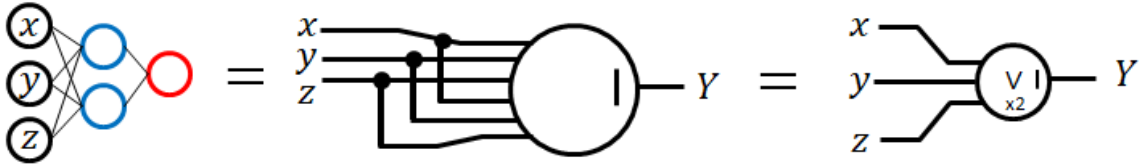

**Fig. 7** Left: network of receptrons for logic disjunction operation for activation conditions in two closed domains (see Fig. 6). Center: Implementation of the network with a multidimensional (6-inputs) receptron capable of solving the function $F = (f_1 \cdot f_2 \cdot f_3) \vee (f_4 \cdot f_5 \cdot f_6)$. Right: symbol proposed for the logic disjunction operation using selective receptrons where "x2" is the number of closed domains and "∨" recalls the symbol of the disjunction operation.

The implementation of the logic disjunction operation can be an effective solution to avoid the catastrophic forgetting problem typical of perceptron based neural networks.

## 6. Conclusions

We have demonstrated that the receptron has intrinsic multi-dimensional selective capabilities for analog inputs due to the input dependent weights, as it is typical of biological neural systems. We propose hardware implementation schemes that use single devices or networks based on the receptron model for multi-dimensional selectivity in both closed (such as cubic or parallelepiped domains) and open domains. Furthermore, receptron-based implementations have been devised for the disjunction operation between closed domains in n-dimensional spaces (for instance, a disjoint pair of cubic domains), highlighting how such implementations can be realized using networks of $m + 1$ receptrons with $n$ inputs or single receptrons with $nxm$ inputs, where $m$ is the number of closed domains. The selective properties of receptrons in closed and open domains and the implementation of logic disjunction operation can be used to avoid the catastrophic forgetting phenomenon in analogy with biological neural networks [23].

We suggest that receptron networks can be used to develop a new class of devices for applications requiring high selectivity and classification capabilities without the burden of complex training typical of classical perceptron-based ANNs. The integration of single receptrons in larger networks could also increase substantially the input capacity and allow the compatibility with many analog inputs. This capability could lead to a substantial improvement of hardware efficiency in edge applications requiring continuous learning and adaptability.